\def\etal{{et al. }}

\def\keV{{\rm keV}}

\documentclass[preprint2]{aastex}
\usepackage{epsfig}
\begin{document}
\shorttitle{GALAXY FORMATION \& THE ICM}
\shortauthors{VOIT \& PONMAN}
\slugcomment{\apj  Letters, 10 Sept 2003}
\title{Signatures of Galaxy Formation in the Intracluster Medium}
\author{G. Mark Voit\altaffilmark{1}$^,$\altaffilmark{2} \&
        Trevor J. Ponman\altaffilmark{3}
         } 
\altaffiltext{1}{Department of Physics and Astronomy,
                 Michigan State University, 
                 East Lansing, MI 48824, 
                 voit@pa.msu.edu}
\altaffiltext{2}{Space Telescope Science Institute}
\altaffiltext{3}{School of Physics and Astronomy,
                 University of Birmingham, 
                 Edgbaston,
                 Birmingham B15 2TT, UK, 
                 tjp@star.sr.bham.ac.uk}

\begin{abstract}
The intergalactic gas in groups and clusters of galaxies
bears the indelible stamp of galaxy formation.  We present
a comparison between observations and simple theoretical
models indicating that radiative cooling governs the entropy 
scale that sets the core radius of the intracluster medium.
Entropy measured at the radius $0.1 r_{200}$ scales as 
$T^{2/3}$, in accord with cooling-threshold models for the
regulation of core entropy.  Cooling of baryons to form
galaxies is likely to lead to feedback, and the signature 
of feedback may appear farther out in the cluster.
Entropy measured at the radius $r_{500}$ in all but
the most massive clusters exceeds the amount that can be
generated by hierarchical accretion.  However, feedback
that smoothes the density distribution of accreting baryons,
perhaps via galactic winds, can boost entropy production
at the accretion shock by a factor $\sim$2-4.  An initial
comparison of entropy at $r_{500}$ to smooth accretion models
shows that smooth accretion is a plausible explanation for
this excess entropy and suggests that baryon accretion 
onto groups was smoother than baryon accretion onto clusters.
\end{abstract}

\keywords{cosmology: theory --- galaxies: clusters: general --- 
galaxies: evolution --- intergalactic medium --- 
X-rays: galaxies: clusters}

\setcounter{footnote}{0}

\section{Introduction}

Clusters and groups of galaxies are interesting
laboratories for studying the cooling and feedback
processes that govern galaxy formation because
these non-gravitational processes have clearly
altered the structure of intracluster and intragroup
media.  Simulations show that the density profiles
of dark-matter halos created by hierarchical
structure formation are nearly self-similar,
with a moderate trend for dark matter to be more
concentrated toward the centers of lower-mass
haloes (e.g., Navarro, Frenk, \& White 1997).  
The baryonic density profiles 
of those halos in purely gravitational simulations
are also nearly self-similar, leading to an X-ray 
luminosity-temperature relation that agrees with
self-similar scaling: $L_X \propto T_X^2$ in the
bremsstrahlung-dominated regime (Muanwong \etal 2001;
Borgani \etal 2001).

Cluster observations showing that $L_X \stackrel 
{\propto} {\sim} T_X^3$ for $T_X \gtrsim 2 \, \keV$
were the first clear indication that galaxy formation 
somehow breaks the self-similarity of the baryonic 
component (e.g., Edge \& Stewart 1991; David \etal 1993;
Markevitch 1998; Arnaud \& Evrard 1999).  This effect shows 
up even more dramatically in the steeper $L_X$-$T_X$
relation for groups (Helsdon \& Ponman 2000) 
and the surprising faintness of the unresolved 
$\sim 1 \, \keV$ background (Pen 1998; Wu, Fabian,
\& Nulsen 2001; Bryan \& Voit 2001).
Recent observations resolving the temperature
structure of clusters have revealed similarity
breaking in the mass-temperature relation as
well.  Self-similar models with $M_\Delta \propto T_X^{3/2}$,
where $M_\Delta$ is the total mass within a sphere whose
mean density is $\Delta$ times the critical density,
agree well with purely gravitational simulations 
(e.g., Evrard, Metzler, \& Navarro 1996; Bryan \& Norman 1998).
However, X-ray observations that resolve the 
temperature structure of clusters now indicate
a slightly steeper slope $M_\Delta \stackrel {\propto}
{\sim} T_X^{1.8}$ and a mass normalization $\sim$1.4
times larger than in simulations without galaxy 
formation (Horner, Mushotzky, \& Scharf 1999;
Nevalainen, Markevitch, \& Forman 2000; Finoguenov,
Reiprich, \& B\"ohringer 2001; Sanderson \etal 2003).

Including galaxy formation in the simulations
generally steepens the $L_X$-$T_X$ relation,
but it is not clear whether the dominant
mechanism responsible for similarity breaking
is cooling or heating.  Simulations that
preheat the intracluster medium by imposing
an entropy floor equivalent to the 100-150$ \, 
\keV \, {\rm cm^2}$ levels observed 
in the cores of groups by Ponman \etal (1999)
produce good agreement with the observed
$L_X$-$T_X$ relation (Bialek, Evrard, \& Mohr 2001).\footnote{
Entropy in this paper is quantified 
in terms of the X-ray observable $K = T n_e^{-2/3}$,
where $n_e$ is the electron density.  The
standard specific entropy scales with the
logarithm of this quantity.}  
Yet, simulations that include radiative cooling without a
compensating heat source seem to agree equally well
with the data, at the expense of an uncomfortably 
large condensed baryon fraction ($\gtrsim 30$\%)
on group scales (Muanwong \etal 2001; Dave, Katz,
\& Weinberg 2002; Tornatore \etal 2003; Kay, Thomas,
\& Theuns 2003).  

The insensitivity of the $L_X$-$T_X$ relation
to the details of galaxy formation can be understood
in terms of the entropy threshold for cooling
within a Hubble time (Voit \& Bryan 2001).
Gas at $\sim 1 \, \keV$ with entropy equivalent to
$\lesssim 100 \, \keV \, {\rm cm^2}$
can radiate away its thermal energy in less than
a Hubble time.  It must either condense or be
heated by feedback to a higher entropy level.
The cooling threshold therefore imprints a physical
entropy scale on a medium that would otherwise be 
nearly scale-free.  A flattening of the entropy distribution
below the cooling threshold, not necessarily an absolute entropy floor,
is all that is needed to produce the observed $L_X$-$T_X$
relation, as long as it forces the cluster's density profile 
to be flatter than $n(r) \propto r^{-3/2}$ where the cooling
time is less than a Hubble time.  Then, the bulk of the cluster's 
luminosity comes from regions where the cooling time is similar 
to a Hubble time, linking the cluster's overall luminosity 
to the cooling threshold (Voit \etal 2002). 
Simulations of cluster formation performed with 
a wide range of feedback efficiencies
support this basic picture, as long as that feedback
heats the affected gas to $\gtrsim 1 \, \keV$ (Kay et 
al. 2003).  In order to distinguish the effects of cooling from
those of heating, more information is needed, and some
of that information can be found in observations that
resolve the entropy profiles of clusters and groups
(e.g., Ponman \etal 2003), which depend in detail
on the interplay between cooling, heating, and 
hierarchical accretion (Voit \etal 2003).

The purpose of this {\em Letter} is to highlight two
trends found in those spatially resolved entropy observations
and to present an initial comparison with theoretical predictions
by way of simple analytical models.
Section~2 focuses on the core entropy measured at $0.1 r_{200}$, 
where $r_{200}$ is the radius at which $\Delta = 200$.
Early measurements of core entropy suggested
that a global entropy floor $\sim 135 \, \keV \,
{\rm cm^2}$ applied to all groups and clusters 
(Ponman \etal 1999; Lloyd-Davies, Ponman, \& Cannon 2000), 
but updated measurements indicate that entropy at
$0.1 r_{200}$ scales as $T^{2/3}$ (Ponman \etal 2003), 
as expected if the cooling threshold indeed determines 
core entropy.  Section~3 focuses on entropy measured
at $r_{500}$, where $\Delta = 500$.  Somewhat unexpectedly,
entropy at these much larger radii also exceeds the predictions
of self-similar models, sometimes by as much as $\sim 10^3
\, \keV \, {\rm cm^2}$ (Finoguenov \etal 2002; Ponman
\etal 2003).  Comparing these measurements to analytical
models for smooth accretion suggests that smoothing of 
the intergalactic medium before it accretes onto a cluster 
or group is a promising mechanism for producing this entropy
excess (Voit \etal 2003; Ponman \etal 2003).
In \S~4 we discuss the implications of these results
and offer a few hypotheses to test with numerical simulations.

\section{Core Entropy and Cooling}

The entropy of intracluster and intragroup media is generated
primarily by shocks owing to mergers and accretion.  Simulations
show that these gravitationally driven processes lead to
an entropy profile $K(r) \stackrel {\propto} {\sim} r^{1.1}$
(Borgani \etal 2001), a scaling that reflects the history of
mass accretion onto the cluster (Tozzi \& Norman 2001; Voit \etal 2003).
Alterations of that profile owing to non-gravitational 
heating and radiative cooling should be most obvious at small
radii, where the gravitationally generated entropy is smallest.
Models involving strong global preheating typically predict
that cluster cores should be nearly isentropic, at the level
of the global entropy floor (Balogh, Babul, \& Patton 1999;
Tozzi \& Norman 2001; Babul \etal 2002).  In models that rely more
on cooling to break self-similarity, the details of the core
entropy profile depend on the interplay between cooling, feedback,
and merging, and perhaps on conduction as well (Voit \etal 2002).  
Regardless of these details, the entropy scale of similarity breaking 
set by cooling should correspond to the cooling threshold $K_c(T) 
\propto T^{1/3} \Lambda^{2/3}$, which leads to $K_c \propto 
T^{2/3}$ when bremsstrahlung dominates the 
cooling function $\Lambda$.

\begin{figure}[t]
\includegraphics[width=3in]{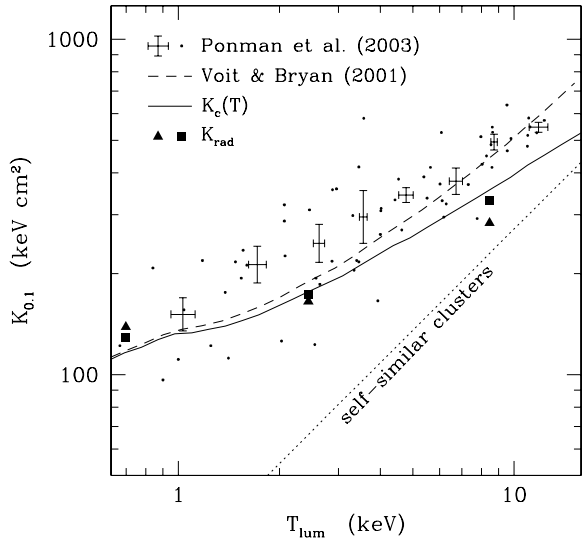}
\caption{ \footnotesize
Relationship between core entropy and the cooling threshold.  
Each point with error bars shows the mean core entropy $K_{0.1}$, 
measured at $0.1 r_{200}$, for eight clusters within a given 
bin of luminosity-weighted temperature $T_{\rm lum}$, and small 
circles show measurements for individual clusters (Ponman \etal 2003).
The dotted line shows a self-similar relation calibrated using 
the median value of $K_{0.1}$ measured in simulation L50+ of 
Bryan \& Voit (2001) which does not include cooling or feedback.
The solid line shows the cooling threshold $K_c(T)$, defined 
to be the entropy at which the cooling time equals 14~Gyr,
assuming the cooling function of Sutherland \& Dopita (1993) 
for 0.3 solar metallicity.  The dashed line shows the entropy
at $0.1r_{200}$ in the model of Voit \& Bryan (2001) when this cooling
function is used.  Solid points without error bars show the total
radiative entropy loss $K_{\rm rad}$ computed for individual halos 
with the accretion histories of Tozzi \& Norman (2001, squares) and 
Voit \etal (2003, triangles).  These simple treatments of cooling
reproduce the slope but slightly underestimate the normalization
of the observed $K_{0.1}(T_{\rm lum})$ relation, which shows no
sign of an absolute entropy floor.
\label{ent_core}}
\end{figure}

Recent observational evidence favors cooling-threshold models.
Figure~\ref{ent_core} shows measurements of core entropy 
$K_{0.1}$ drawn from a compilation of cluster observations.
Ponman \etal (2003) collected data from 64 clusters and groups
and presented the mean entropy levels within temperature bins
containing eight clusters each.  Binning the data averages 
out the cluster-to-cluster variations, which can range
up to a factor of three at a given temperature (e.g., Mushotzky
\etal 2003).  These measurements sit well above the dotted 
line showing a self-similar relation calibrated to match
the purely gravitational simulations of Bryan \& 
Voit (2001).  However, the measurements do not level off 
at a well-defined entropy floor.  Instead, they trace a
power-law relation $K_{0.1} \propto T^{0.65}$, nearly
identical to the scaling of the cooling threshold
(Ponman \etal 2003).

Entropy profiles of individual clusters and groups likewise 
show no evidence for an absolute entropy floor.  The
cores of groups observed with {\em XMM-Newton} are
not isentropic but instead have entropy profiles
similar to those of clusters (Mushotzky \etal 2003;
Pratt \& Arnaud 2003).  In fact, scaling those profiles
according to $K \propto T^{2/3}$ produces a much closer
match to massive-cluster profiles than the 
expected self-similar scaling $K \propto T$
(Pratt \& Arnaud 2003; Ponman \etal 2003).  Furthermore,
the core entropy levels required to explain the $L_X$-$T_X$
relation with isentropic models are far higher than those
observed (Voit \etal 2003).

Simplistic models involving the cooling threshold come 
close to matching the $K_{0.1}$-$T_{\rm lum}$ relation, but 
the match is not exact.  The solid line in Figure~\ref{ent_core} 
shows the cooling threshold $K_c(T_{\rm lum})$, defined 
to be the entropy level at which the cooling time equals 14~Gyr,
according to the cooling function for 0.3
solar metallicity from Sutherland \& Dopita (1993).  This
threshold lies $\sim$25\% below the binned data points with
a very similar slope.  The dashed line shows $K_{0.1}$
computed from $K_c$ using the prescription of Voit \&
Bryan (2001; see also Wu \& Xue 2002a,b).  In these models the 
self-similar entropy distribution of unmodified intracluster
gas is simply truncated below $K_c$, and the gas is allowed
to relax back into hydrostatic equilibrium.  These 
truncated models agree better with the hotter 
clusters but not do not provide much enhancement over 
$K_c$ near $\sim$1~keV because this simplistic prescription 
leads to nearly isentropic cores in the low-mass clusters.
Treating cooling more realistically produces an entropy
gradient within $0.1 r_{200}$ that declines to zero entropy 
at the origin, in the absence of feedback and conduction 
(Voit \etal 2002).

A more detailed computation of $K_c$ accounting for 
time dependence of the gas temperature does not change
this basic result.  One can compute the total entropy
loss $K_{\rm rad}$ owing to radiative cooling assuming
that the gas temperature at a given time equals the 
temperature of the most massive progenitor halo at that 
time (see Voit \etal 2003).  The squares and triangles in
Figure~\ref{ent_core} show that $K_{\rm rad}$ differs
little from $K_c$ for a typical merger history (see
also Oh \& Benson 2003).

Taken as whole, Figure~\ref{ent_core} strongly suggests
that radiative cooling regulates the entropy scale that defines
the cores of clusters.  However, the large dispersion of the
unbinned data points indicates that cooling acts in concert
with other processes.  Much of that dispersion may arise
from the stochastic nature of mergers, which produce 
significant entropy differences in clusters of similar
temperature (see Figure 5 of Voit \etal 2003).  
Differing feedback histories may also be responsible
for some of this dispersion.

\section{Outer Entropy and Smoothing}

Elevated entropy levels in cluster cores were once thought
to be evidence for non-gravitational energy input into the
intracluster medium, but the close link between core entropy
and the cooling threshold now renders that interpretation of core
entropy somewhat questionable.  Nevertheless, feedback from galaxy 
formation still seems necessary to limit the fraction of
condensed baryons in clusters.  In models of hierarchical
structure formation, a large percentage of the plasma that 
will become the intracluster medium has a cooling time less than
a Hubble time at $z \sim 2$-3 (e.g., Voit \etal 2002).  
Because only $\sim$5-10\% of baryons are observed to be 
in stars (e.g., Balogh \etal 2001), 
feedback from supernovae and active galactic nuclei presumably
intervenes to prevent most of this intergalactic gas from 
condensing.  Here we present evidence suggesting that this
early episode of cooling and feedback may couple with
subsequent merging and accretion to produce an observable 
entropy signature in the outskirts of clusters owing to the
smoothing effect of galactic feedback on the intergalactic medium.

Hierarchical merging in the absence of cooling and feedback
tends to produce a nearly self-similar entropy distribution
(Voit \etal 2003).  The characteristic entropy associated
with the baryons in a halo of mass $M_{200}$ is 
$K_{200} \equiv T_{200} \bar{n}_e^{-2/3}$, where 
$T_{200} = G M_{200} \mu m_p / 2 r_{200}$  is the characteristic 
temperature of the halo and $\bar{n}_e \approx 1.6 \times 10^{-4} \, 
{\rm cm}^{-3}$ is $200 \Omega_M^{-1}$ times the mean electron 
density of the universe.  According to the simulations of
Bryan \& Voit (2001), the maximum value of intracluster
entropy is typically $\approx 0.8 K_{200}$ without galaxy formation,
and the median entropy value at the scale radius $r_{500}$
is $\approx 0.4 K_{200}$.  These scalings do not appear
to depend significantly on halo mass.

\begin{figure}[t]
\includegraphics[width=3in]{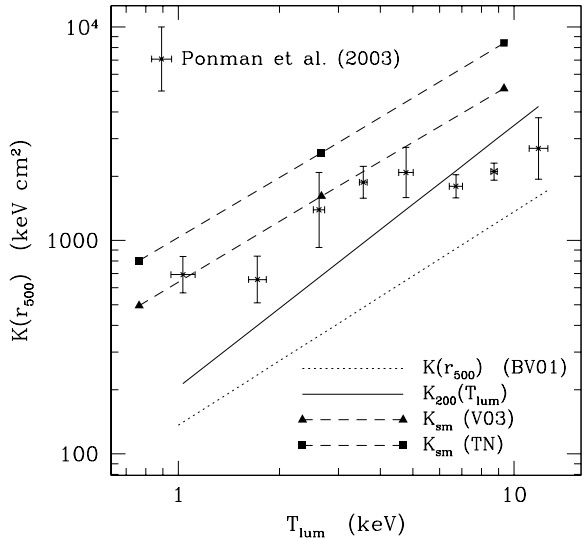}
\caption{ \footnotesize
Entropy at $r_{500}$ as a function of luminosity-weighted
cluster temperature.  Each point with error bars shows the mean
value of $K(r_{500})$ implied by the density and temperature profiles
of eight clusters within that temperature bin (Ponman \etal 2003).
The dotted line shows the predicted entropy at $r_{500}$ for self-similar
clusters calibrated with simulation L50+ of Bryan \& Voit (2001).
The solid line shows the entropy scale $K_{200}$ computed from 
$T_{\rm lum}$ using the $M_{200}$-$T_{\rm lum}$ relation of Sanderson 
\etal (2003), which is the maximum amount of intracluster entropy expected 
from hierarchical accrection in the absence of non-gravitational processes
(see Voit \etal 2003).  
Dashed lines connect the maximum entropy values produced by smooth accretion
onto individual halos, according the accretion histories of Voit \etal
(2003, triangles) and Tozzi \& Norman (2001, squares).  The data
suggest a transition from hierarchical accretion in the most massive 
clusters to smooth accretion on group scales.
\label{ent_excess}}
\end{figure}

In contrast, observations imply a considerably higher entropy value 
at $r_{500}$ with a clear dependence on halo mass (Finoguenov
\etal 2002; Ponman \etal 2003).  Figure~\ref{ent_excess} shows
mean values of $K(r_{500})$ inferred by Ponman 
\etal (2003) from the same data set as in Figure~\ref{ent_core}.
Notice that all of the binned data points lie above the dotted line showing
the value of $K(r_{500})$  predicted by self-similar models
calibrated with the simulations of Bryan \& Voit (2001), indicating
that galaxy formation affects the entire entropy profile of
a cluster, not just the core.  
The observed entropy boost is quite large.  A solid line showing 
the value of $K_{200}$ inferred from $T_{\rm lum}$ and the observed 
$M_{200}$-$T_{\rm lum}$ relation of Sanderson \etal (2003) indicates
the maximum entropy expected from pure hierarchical accretion.  
All the data points below $\sim 5 \, \keV$ lie above this line 
by hundreds of keV~cm$^2$, representing an entropy excess that 
is very difficult to produce with non-gravitational energy 
injection alone (Ponman \etal 2003).

Voit \etal (2003) and Ponman \etal (2003) independently
suggested that this excess entropy arises because feedback
has smoothed the baryonic matter that accretes onto a cluster,
thereby reducing the characteristic density of gas that passes 
through the accretion shock.  Smoothing boosts intracluster 
entropy because the entropy of strongly shocked gas is 
$\propto v_s^2 \rho_1^{-2/3}$, where $v_s$ is the shock velocity 
and $\rho_1$ is the preshock baryon density.   The scale 
of $v_s$ is determined by dark-matter dynamics that are 
virtually unaffected by cooling and feedback.  Thus, feedback 
raises the mass-weighted average value of $v_s^2 \rho_1^{-2/3}$ because 
it reduces the mass-weighted average value of $\rho_1$ 
(Voit \etal 2003).  Notice that a modest amount of preshock 
entropy can be strongly amplified at the accretion
shock of a deep potential well if that preshock entropy is
sufficient to drive a large proportion of baryons out of the
shallower potential wells of the accreting dark matter. 
Ponman \etal (2003) envisioned this smoothing as a 
thickening of the filaments through which
matter accretes.  Voit \etal (2003) modeled this entropy increase
by developing analytical models for entropy production during
smooth accretion.

The models of Voit \etal (2003) show that smooth, spherically
symmetric accretion produces an entropy profile whose shape
is similar to that found in simulations without galaxy formation
but whose normalization is $\sim$2 times higher in massive clusters
and up to $\sim$4 times higher in groups.  The magnitude of this
entropy excess depends on the mass-accretion rate---objects that
accrete most of their mass late in time have a higher baryon
density entering the accretion shock and therefore less of an
entropy excess.  The dashed lines in Figure~\ref{ent_excess}
indicate the maximum amount of entropy produced by smooth
accretion, assuming the accretion histories of Tozzi \& Norman
(2001; squares) and Voit \etal (2003; triangles) for halo
masses of $10^{13} \, h^{-1} \, M_\odot$, $10^{14} \, h^{-1} \, M_\odot$,
and $10^{15} \, h^{-1} \, M_\odot$.

This initial comparison between analytical models for
smooth accretion and observed entropy levels at $r_{500}$ suggest
that smoothed accretion is a plausible explanation for 
these entropy excesses because the data points in
Figure~\ref{ent_excess} approach but do not exceed 
the dashed lines.  Furthermore, the
trend of the data points with $T_{\rm lum}$ suggests that
accretion of baryons onto groups is smoother than accretion
of baryons onto clusters.  This apparent transition from
nearly smooth accretion at $\lesssim 1 \, \keV$ to purely
hierarchical accretion at $\gtrsim 10 \, \keV$ may also
explain why the entropy profiles of groups are similar in 
shape to those of clusters but have shifted normalizations 
(e.g., Pratt \& Arnaud 2003).  Curiously, scaling the 
profiles by the same $T^{2/3}$ factor found in the cores
brings them into agreement, suggesting that the cooling
threshold plays a role in setting the smoothing scale.
However, it is not immediately clear how cooling and feedback
would couple with accretion to produce this scaling.

\section{Discussion}

The entropy of baryons in both the cores and outskirts
of clusters disagrees with simulations that 
do not include galaxy formation.  Thus, the entropy 
profiles of these objects contain valuable information
about the impact of galaxy formation on the intergalactic
medium and the role of feedback in regulating global
star formation.  The scaling of core entropy 
($K_{0.1} \propto T_{\rm lum}^{2/3}$) suggests that 
it is governed by radiative cooling  (Ponman \etal 2003).  
However, feedback is likely to limit the fraction of
baryons that eventually cool and form stars.

If feedback is strong enough to eject a substantial
percentage of baryons from low-mass halos, then this
smoothing of the baryon distribution will affect
entropy production during subsequent accretion and
mergers.  Comparing smooth-accretion models with the 
available data (Figure~\ref{ent_excess}) suggests 
that a large percentage of the matter accreting onto 
groups and low-mass clusters must be smoothed in order 
to explain the observed entropy excess at $r_{500}$.  
If this explanation is correct, then smoothing may 
also have an observable impact on the X-ray substructure 
in and around low-mass clusters and the associated 
filaments.  Measuring how feedback affects the level 
of X-ray substructure in numerically simulated groups and
clusters would help observers to quantify the level
of intergalactic smoothing. 

Only in the highest-mass clusters do entropy profiles 
approach the self-similar form predicted by simulations
without cooling and feedback.  However, these clusters
presumably formed through mergers of lower-mass objects
in which cooling and feedback had already broken 
self-similarity.  We speculate that hierarchical merging
restores similarity because the self-similar entropy
profile behaves as a dynamical attractor:  As structure
grows, the deepening potential wells are increasingly
able to overcome baryon smoothing produced by early feedback.
Accretion onto massive clusters is therefore lumpier
than accretion onto groups, and the resulting entropy 
profile approaches self-similarity when the accreting
lumps of dark matter are able to retain the majority
of their associated baryons. 

Many questions about the link between galaxy formation
and the intracluster medium remain unanswered.  Among
the most crucial is the amount of energy injection
implied by the entropy profiles of clusters.  Numerical
simulations that track entropy production in low-mass
halos would be very helpful because smoothing allows 
a small amount of entropy injection to be amplified 
with each merger event during the formation of a group
or cluster.  High resolution simulations will therefore 
be needed to assess how early feedback propagates through 
the entire merger hierarchy and to characterize how
the hottest clusters eventually approach self-similarity.

\vspace*{1em}
The authors thank Greg Bryan for sharing his insights into
simulated clusters and Alistair Sanderson for his contributions
to the analysis of the observational data.
GMV received partial support from NASA through
grant NAG5-3257.  


\end{document}